%% file: paper_arxiv.tex
\documentclass[preprint,aps,prl,groupedaddress]{revtex4-1}

\usepackage{hyperref} 
\hypersetup{
colorlinks=true,
			citecolor = blue,
            linkcolor=blue,
            urlcolor=blue,
    }

\usepackage{graphicx} 
\usepackage{epstopdf} 
\usepackage{float} 

\begin{document}

\title{Probing the Nanoskyrmion Lattice on Fe/Ir(111) with Magnetic Exchange Force Microscopy}
\author{Josef Grenz}
\affiliation{Department of Physics, University of Hamburg, Jungiusstra{\ss}e 11A, D-20355 Hamburg}
\author{Arne K\"ohler}
\affiliation{Department of Physics, University of Hamburg, Jungiusstra{\ss}e 11A, D-20355 Hamburg}
\author{Alexander Schwarz}
\affiliation{Department of Physics, University of Hamburg, Jungiusstra{\ss}e 11A, D-20355 Hamburg}
\email{aschwarz@physnet.uni-hamburg.de}
\author{Roland Wiesendanger}
\affiliation{Department of Physics, University of Hamburg, Jungiusstra{\ss}e 11A, D-20355 Hamburg}

\date{\today}

\begin{abstract}
We demonstrate that the magnetic nanoskyrmion lattice on the Fe monolayer on Ir(111) and the positions of the Fe atoms can be resolved simultaneously using magnetic exchange force microscopy. Thus, the relation between magnetic and atomic structure can be determined straightforwardly by evaluating the Fourier transformation of the real space image data. We further show that the magnetic contrast can be mapped on a Heisenberg-like magnetic interaction between tip and sample spins. Since our imaging technique is based on measuring forces, our observation paves the way to study skyrmions or other complex spin textures on insulating sample systems with atomic resolution.
\end{abstract}

\maketitle

Magnetic skyrmions are quasiparticles with a non-trivial spin texture \cite{bogdanov1994}, that features an integer non-zero winding number, and are stabilized by the Dzyaloshinsky-Moriya interaction (DMI) \cite{dzyaloshinsky1958,moriya1960}.
Recently, they have been observed in bulk materials \cite{Muehlbauer915,Yu2010,seki2012} as well as in ultrathin films \cite{vonbergmann2007,heinze2011,yu2011,Romming2013,Fert2016}. Magnetic skyrmions can be manipulated in various ways and thus are promising candidates for magnetic data storage applications or other spintronic devices and hence attract a lot of attention \cite{Kiselev2011,nagaosa2013,fert2013,hsu2017}. 
Compared to magnetic bubbles \cite{ODell1986}, which are stabilized by the long-range magnetostatic dipolar interaction, magnetic skyrmions are usually much smaller, especially in ultrathin films where the interfacial DMI can be of significant strength \cite{heinze2011,wiesendanger2016}. 

Hitherto, the smallest magnetic skyrmions (nanoskyrmions) have been observed using spin-polarized scanning tunneling microscopy (SP-STM) \cite{wiesendanger2009} in the Fe monolayer on Ir(111), where they form a square lattice with a periodicity of about 1 nm \cite{vonbergmann2006,vonbergmann2007,heinze2011}. Unlike in other skyrmion systems, this lattice is the magnetic ground state and not a metastable phase that only exists at certain temperatures and ranges of magnetic field. Up to now, high-resolution studies of non-collinear spin textures down to the atomic scale could only be performed with SP-STM and thus were limited to conductive sample systems. However, recently magnetic skyrmions have been discovered in insulating oxide systems as well \cite{seki2012}, and therefore techniques capable of resolving non-collinear spin states with atomic resolution on non-conductive samples are of significant interest.

Here, we report magnetic exchange force microscopy (MExFM) results obtained on the well-known Fe monolayer on Ir(111) as model-type nanoskyrmion system. This atomic force microscopy (AFM) based method has already proven to resolve the spin structure of a collinear antiferromagnetic bulk insulator, namely NiO(001) \cite{kai2007}. Our findings demonstrate that MExFM can be applied in future search for skyrmions in novel types of insulating sample systems. Particularly, we find that MExFM allows the atomic and magnetic structure of the nanoskyrmion lattice to be resolved simultaneously. Therefore, determining the relation between atomic and magnetic structure is much simpler than for SP-STM data \cite{vonbergmann2006,vonbergmann2007,heinze2011}. Even in a recent combined SP-STM/MExFM study this could not be achieved \cite{hauptmann2017}. In addition, we utilize  the non-collinear spin texture of this sample system to verify experimentally that MExFM can indeed be interpreted using a simple Heisenberg exchange interaction between spins at the tip apex and at the surface \cite{kai2007}.

The MExFM experiments were carried out at low-temperatures ($8.2\text{ K}$) in ultra-high vacuum (UHV) with a home-built atomic force microscope \cite{liebmann2002}. Additionally, a magnetic field of up to $5 \text{ T}$ perpendicular to the sample surface can be applied. All measurements were performed in the non-contact regime using the frequency modulation technique \cite{albrecht1991}.  In this mode of operation, the cantilever is self-excited with a shaker piezo and oscillates with its resonance frequency $f$ while the amplitude $A_0$ is kept constant by a feedback circuit that resupplies any dissipated energy by adjusting the excitation amplitude $a_{\text{exc}}$ accordingly. Forces between the tip and sample shift the actual resonance frequency $f$ by $\Delta f = f-f_0$, where $f_0$ is the eigenfrequency of the cantilever without any tip-sample interaction. To image the sample surface the tip is scanned line by line across the $(x,y)$ sample surface while $\Delta f$ is kept constant. Thus, the $z$-feedback is constantly adjusting the tip-sample distance $z(x,y)$, which corresponds to the surface topography. The contact potential difference (CPD) between the tip and sample was determined by recording $\Delta f(U_{\text{bias}})$-curves and compensated by applying an appropriate compensation voltage $U_{\text{bias}} = U_{\text{CPD}}$ to minimize long-range electrostatic interactions. As it has been described previously \cite{teobaldi2011}, these curves were also used to characterize the tips.

MExFM imaging at close proximity was performed with constant height $z$ while additionally keeping $a_{\text{exc}}$ constant and recording $\Delta f\left(x,y\right)$. Thereby, drift and creep related height changes, which are present if the $z$-feedback is simply switched off and are particularly troublesome at small tip-sample separations, could be avoided. Furthermore, since the dissipation and thus $a_{\text{exc}}\left(z\right)$ usually increases monotonically with $z$, this mode of operation is less prone to tip crashes at small tip-sample separations compared to imaging at a constant $\Delta f$. We made sure that no atomic or magnetic site dependent dissipation is present, because otherwise cross-talk between the dissipation channel $a_{\text{exc}}\left( x,y \right)$ and $\Delta f\left(x,y\right)$ prevents any straightforward image interpretation. 

To prepare a clean tip without localized charges, supersharp Si cantilevers ($f_0 \approx 150 \text{ kHz}$, spring constant $k_c \approx 147 \text{ Nm}^{-1}$ and $2 \text{ nm}$ nominal tip radius \footnote{NANOSENSORS\texttrademark . Rue Jaquet-Droz 1. CH-2002 Neuch\^{a}tel, Switzerland.}) were degassed and coated afterwards with a few nm of Ti. In order to obtain a magnetic-sensitive tip, a few nm of Fe were evaporated onto the tip as well. To align the tip magnetization to a well defined out-of-plane direction, the external magnetic field oriented perpendicular to the sample surface plane was applied.

\begin{figure}  
        \includegraphics[width=0.5\textwidth]{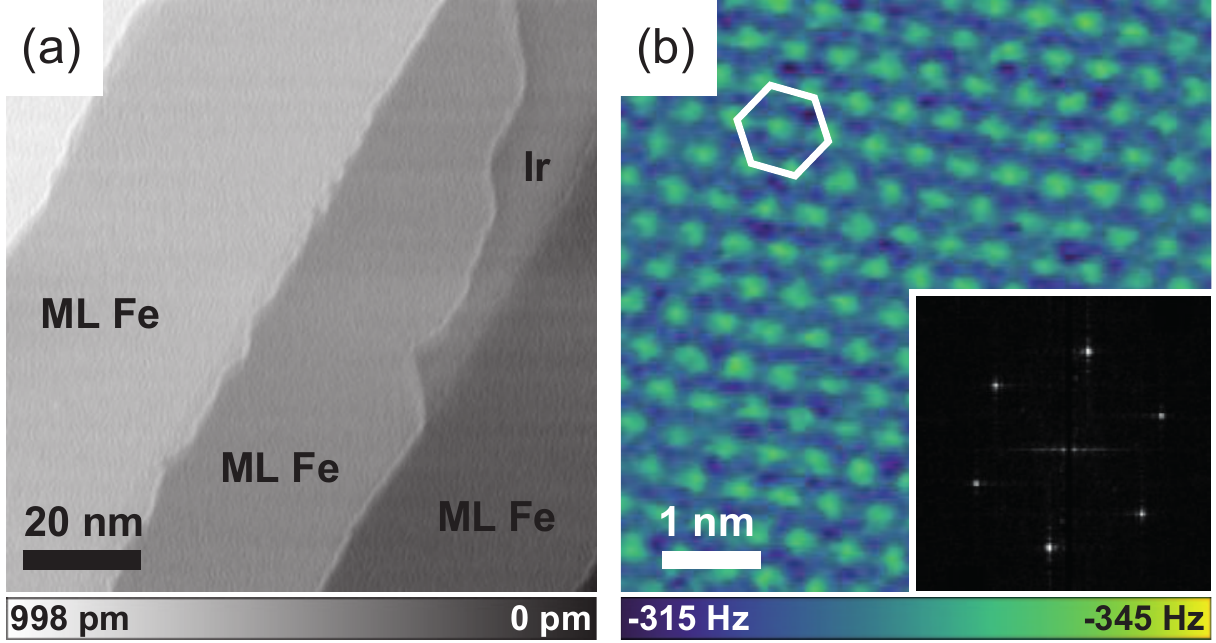}
    \caption[overview]{(a) Overview NC-AFM image of the Fe/Ir(111) sample system with a coverage of 0.7 ML. Parameters: $\Delta f = -12.0 \text{ Hz}$, $A = 2.3 \text{ nm}$, $k_c = 147.1 \text{ Nm}^{-1}$, $U_{\text{bias}} = +0,1 \text{ V}$. (b)
	Atomic resolution image the Fe ML in constant-height mode recorded with a non-magnetic tip. The hexagonal lattice demonstrates that the Fe ML grows pseudomorphic on Ir(111). The inset shows the 2D-FT of the image data with the six spots corresponding to the hexagonal atomic lattice. Parameters: $A = 1.2 \text{ nm}$, $k_c = 148.3 \text{ Nm}^{-1}$, $U_{\text{bias}} = +0.9 \text{ V}$.}
    \label{fig:sample}
\end{figure}

The Fe/Ir(111) sample system was prepared by cleaning the single crystal substrate using cycles of oxygen annealing ($5.0\cdot 10^{-8} \text{ mbar}$, heating stepwise to $1440\text{ K}$) and subsequent Ar$^+$ ion sputtering. As a final step, the Ir crystal was flashed at $1440\text{ K}$ for $60\text{ s}$. Thereafter, Fe was deposited by thermal evaporation onto the substrate held at an elevated temperature of $\sim 440\text{ K}$ while the background pressure was at about $1.5\cdot 10^{-10}\text{ mbar}$. The resulting surface with an Fe coverage of $0.7$~ML is shown in Fig. \ref{fig:sample}(a). Terraces with a width of $\left(50-80\right) \text{ nm}$ are covered by single layer Fe while small areas exist where the bare Ir(111) surface is exposed. In \ref{fig:sample}(b) the atomic structure of the Fe ML is resolved with a non-magnetic tip. The hexagonal structure of the Fe lattice can be clearly identified in the image data and its two-dimensional Fourier transform (2D-FT, see inset). Note that Fe usually crystallizes in the bcc structure but is forced into this unusual hexagonal lattice by its pseudomorphic growth on the Ir(111) substrate.

\begin{figure*} 
        \includegraphics[width=1.0\textwidth]{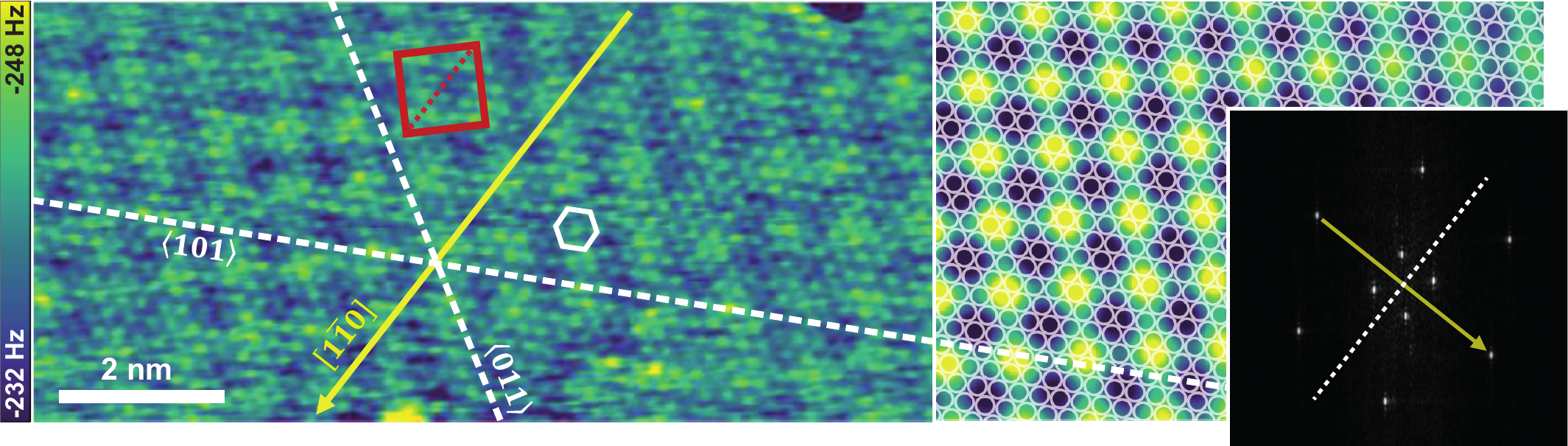}
    \caption[skyrmions]{(left) MExFM image of the Fe ML on Ir(111) recorded with a Fe-coated tip in constant-height mode. An applied magnetic field of $B= 4 \text{ T}$ ensures an out-of-plane magnetized tip. In addition to the hexagonal atomic structure (white hexagon) the magnetic structure of the nanoskyrmion lattice (red square) is visible as well. The dotted lines indicate the principal crystallographic directions of the substrate. (right) Overlaid Fourier-filtered image generated by taking only the atomic and magnetic peaks of the 2D-FT. The atomic lattice has been reproduced by a hexagonal arrangement of white circles. 
    (inset) The 2D-FT of the raw image data clearly reveals that atomic and magnetic structure are incommensurate. Parameters: $A = 0.8 \text{ nm}$, $k_c = 147.2 \text{ Nm}^{-1}$, $U_{\text bias} = +1.1 \text{ V}, B = 4\text{ T}$.}
    \label{fig:lattice}
\end{figure*}

Utilizing a magnetically sensitive tip an additional contrast pattern emerges. Figure \ref{fig:lattice} (left) shows the hexagonal arrangement of atoms and a superimposed larger square lattice. This is the characteristic pattern expected for the nanoskyrmion state. Note that SP-STM images recorded on this particular sample system usually only reveal the magnetic but not the atomic structure \cite{vonbergmann2006,vonbergmann2007,heinze2011}. Taking advantage of the tunneling anisotropic magnetoresistance (TAMR) effect \cite{bode2002}, it has been possible to resolve the atomic lattice together with a periodic structure, that is intimately related to the spin texture \cite{heinze2011}. However, since this contrast mechanism is related to a modulation of the local density of states due to spin-orbit coupling, it is not possible to distinguish between magnetic moments of the same magnitude pointing in opposite directions \cite{heinze2011,hanneken2015}. In contrast, MExFM image data on the Fe/Ir(111) system can be interpreted in a more direct manner.

To further evaluate the details of the contrast pattern, we analyze the 2D-FT of the image data visible in the inset of Fig. \ref{fig:lattice}. Atomic and magnetic structures are represented by outer spots arranged in a hexagon and inner spots arranged in a square, respectively. Assuming pseudomorphic growth, the outer peaks correspond to the lattice constant $a_{\text{hex}}= 384 \text{ pm}$ of the Ir(111) surface \cite{vonbergmann2007}. Thus, the lattice constant of the square lattice can be determined to be $a_{\text{sq}}= \left(902\pm18\right) \text{ pm}$. Clearly, the atomic and magnetic lattices are incommensurate. The [11]-direction of the square lattice is parallel to the closed-packed [1$\bar{1}$0]-direction of the substrate, while the [10]-direction ([01]-direction) is rotated by $30^\circ$ relative to the [101]-direction ([011]-direction) of the substrate. Note that due to the symmetry of the substrate, three rotational domains of the magnetic structure exist (see below). 

The incommensurability can also be seen in the overlaid Fourier-filtered image in the right part of Fig. \ref{fig:lattice}. The image was generated by inverse transformation of the 2D-FT by using only areas around the peaks. The positions of the surface atoms are indicated by white circles, while the magnetic superstructure is shown as a colored texture. Clearly, the maxima of the square lattice are not always on top of an atom.

\begin{figure} 
        \includegraphics[width=0.485\textwidth]{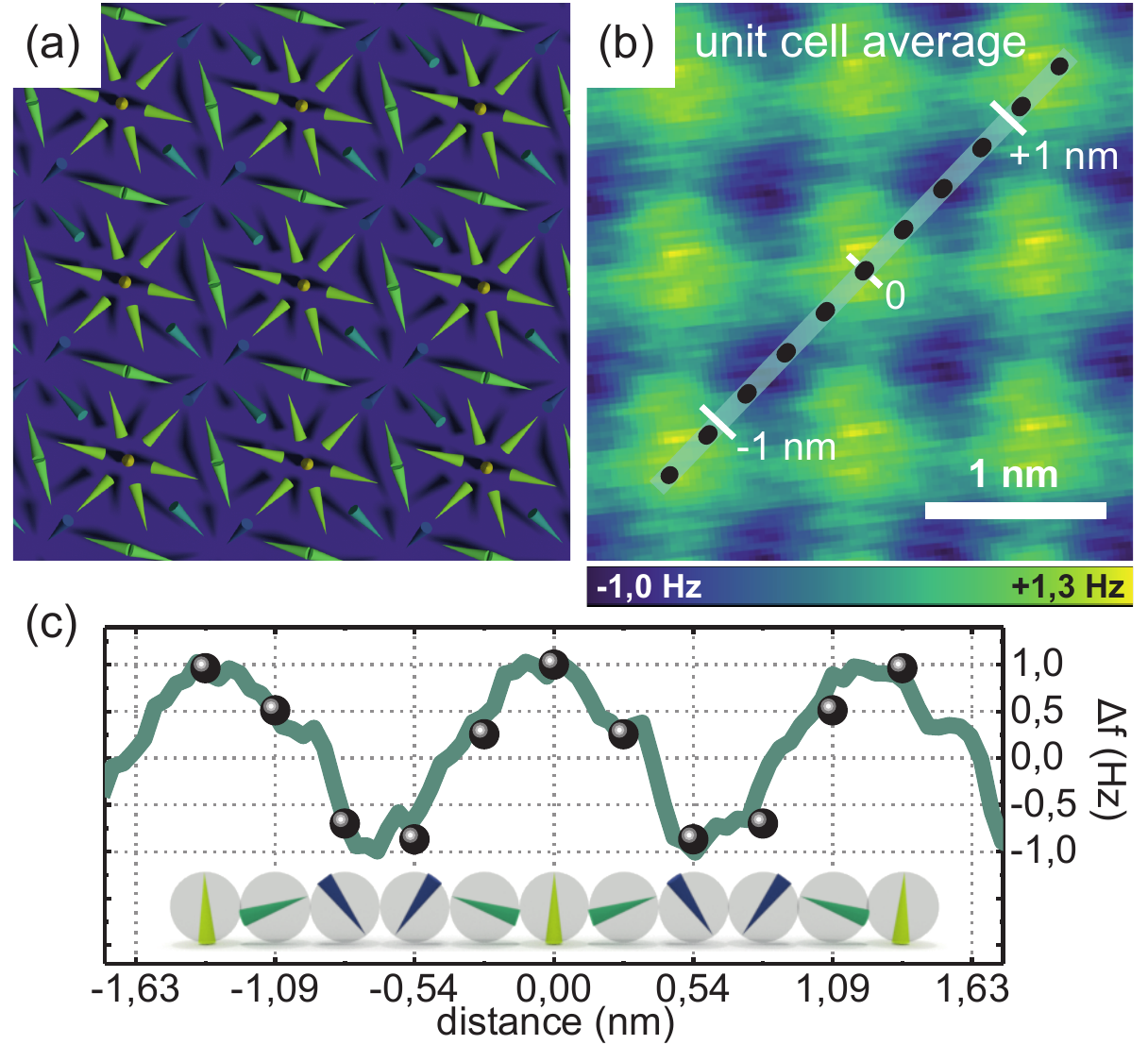}
    \caption[skyrmions]{(a) Commensurate approximation of the spin texture on top of the hexagonal arrangement of atoms that is closest to the real incommensurate nanoskyrmion lattice (adapted from Ref. \cite{heinze2011}). The polar and azimuthal angles of the cones represent the orientation of the canted spins with respect to the surface plane. (b) Magnetic contrast obtained after removing the atomic contribution using Fourier-filtering and subsequent unit cell averaging. The image is tiled from several unit cells to match the size of the model in (a). The line indicates the [11]-direction of the square lattice, which is parallel to [1$\bar{1}$0]-direction of the substrate. Black dots mark the positions of the Fe atoms. (c) Line section as indicated in (b). The black circles represent the expected behavior of the magnetic signal according to the Heisenberg model assuming an out-of-plane magnetized tip. Below the curve, the cones in the spheres indicate the varying angle between spin and surface plane.}
    \label{fig:contrast}
\end{figure}

Comparing the peak heights in the 2D-FT image we find that the ratio between atomic and magnetic signal is close to $1$, i.e., $0.83$ for this particular data set. Note that this ratio is distance- and tip-dependent. However, we find a similar ratio for the antiferromagnetic ML of Fe in W(001) \cite{schmidt2009}, also an itinerant system, while on the antiferromagnetic bulk insulator NiO(001) the magnetic contrast is about 3 times weaker than the chemical contrast \cite{kai2007}. Additionally, we used the magnetic contrast pattern of the nanoskyrmion lattice to verify the previously made assumption in theoretical studies of MExFM contrast formation that the short-range magnetic exchange interaction between tip and sample spins $\vec{S}_t$ and $\vec{S}_s$ can be described by a simple Heisenberg model, i.e., $H = J \left(\vec{S}_t \cdot \vec{S}_s\right)$ with $J$ being the coupling constant \cite{Momida2005,kai2007,schmidt2009,pielmeier2013}. To test this, we assume a commensurate skyrmionic spin texture for the sample as used in Ref. \cite{heinze2011}. It represents a very good approximation for the actual incommensurate spin texture and is displayed in Fig. \ref{fig:contrast}(a). Image \ref{fig:contrast}(b) shows pure magnetic contrast. It was obtained by removing the atomic contrast from the data shown in Fig. \ref{fig:lattice} using Fourier filtering and applying unit cell averaging afterwards. To generate the same image size as in \ref{fig:contrast}(a) several magnetic unit cells were tiled together. In \ref{fig:contrast}(c) a line section along the [11]-direction of the square lattice is plotted. The positions of the atoms and their perpendicular spin orientation relative to the surface plane are sketched below the line section.
The details of the spin configuration at the tip apex are unknown, but due to the short-range nature of the magnetic exchange interaction we assume that the tip can be represented by a single spin that is aligned along the external magnetic field, i.e., perpendicular to the sample surface. At each Fe atom along the [11]-direction we calculated the scalar product between its spin and the spin of tip apex, respectively. The expected relative magnetic signal at the positions of the individual Fe atoms agrees very well with the experimental corrugation, if we multiply all calculated data points with a single scaling factor. Since $J$ as well as $|S_t|$ are unknown, this scaling is required. They could be determined using density functional theory based calculations of the combined tip-sample system as performed in Ref. \cite{schmidt2009} for the much simpler collinear Fe ML on W(001) system. Such calculations are beyond the scope of our work presented here.

\begin{figure} 
	\centering
       \includegraphics[width=0.5\textwidth]{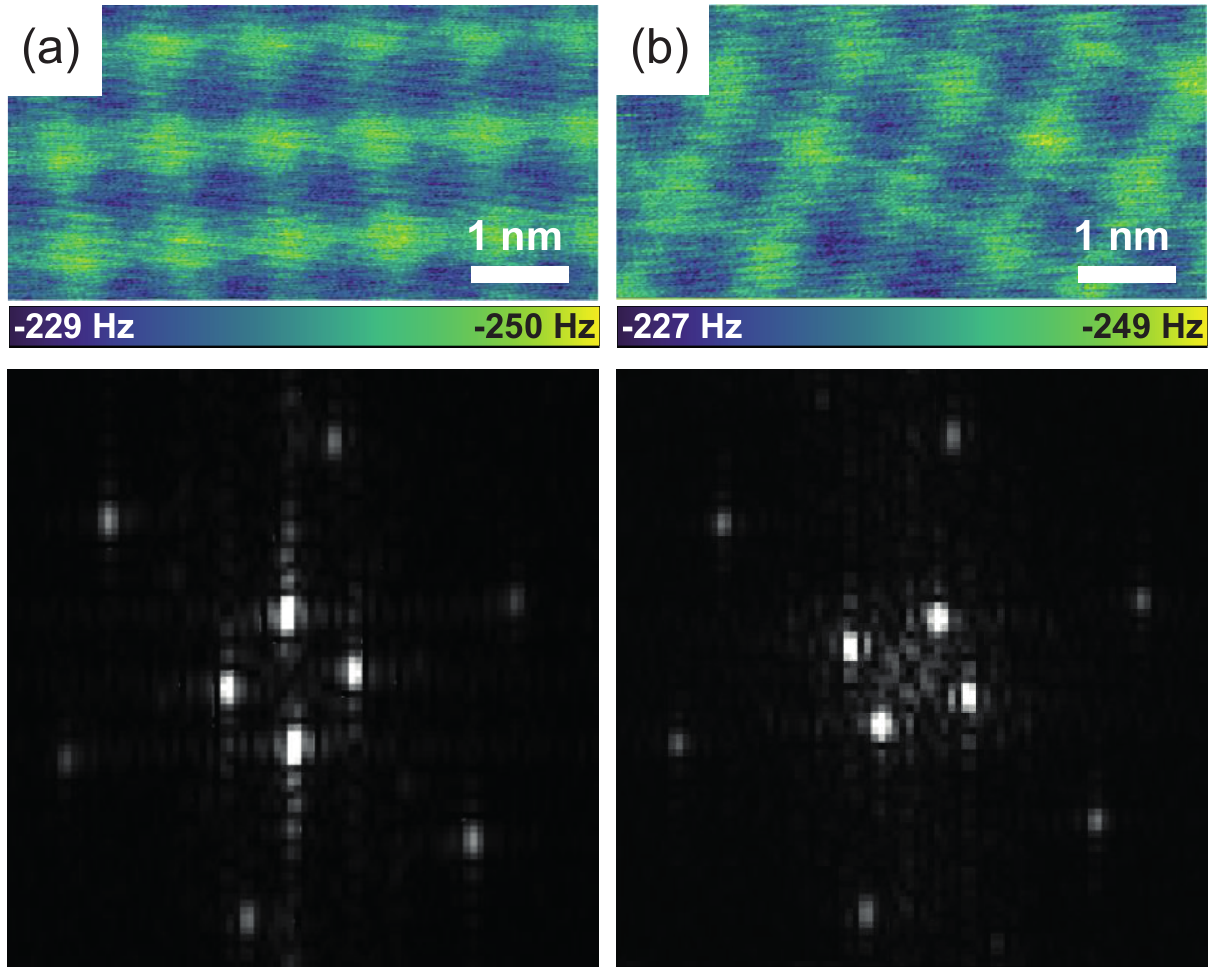}
  \caption[domains]{MExFM images (a) and (b) show two rotational domains of the skyrmion lattice. They were recorded in close proximity to each other using the same constant-height mode imaging parameters. As visible in the corresponding 2D-FT displayed below the images, only the magnetic contrast pattern is rotated. Parameters: $B = 3\text{ T}$, $A = 0.5 \text{ nm}$, $k_c = 147.1 \text{ Nm}^{-1}$, $U_{\text{bias}} = +0.1 \text{ V}$.}
    \label{fig:domains}
\end{figure}

As mentioned above, due to the threefold symmetry of the substrate, three equivalent rotational domains of the skyrmion lattice occur on the surface. In Fig. \ref{fig:domains} two different orientations are displayed. They have been observed with the same tip and with identical imaging parameters in close vicinity to each other. Note that the atomic lattice can hardly be perceived in both images. However, the 2D-FT of the image data reveals the characteristic six spots. Clearly, they are at identical positions in both cases, while the four magnetic spots are rotated by $60^\circ$ against each other.

In conclusion, our findings demonstrate that MExFM is a new versatile tool to investigate skyrmions (or any other complex spin texture) down to the atomic scale by the measurement of forces only. Thus, it will become possible to specifically search for skyrmions in insulating thin films, which cannot be studied with STM-based methods.

\begin{acknowledgments}
Financial support from the DFG in the framework of SFB668 (TP A5) and the European Union via the Horizon 2020 research and innovation program under grant agreement No. 665095 (MagicSky) is gratefully acknowledged.
\end{acknowledgments}

\input{paper.bbl}

\end{document}

%% file: paper.bbl
%